\documentclass[12pt,a4paper]{article}
\usepackage{amsmath}
\usepackage{latexsym}
\makeatletter
\def\rddots{\mathinner{\mkern1mu\raise\p@%
    \vbox{\kern7\p@\hbox{.}}\mkern2mu%
    \raise4\p@\hbox{.}\mkern2mu\raise7\p@\hbox{.}\mkern1mu}}
\makeatother
\begin{document}

\title{\sl Beyond Gaussian : A Comment}
\author{
  Kazuyuki FUJII
  \thanks{E-mail address : fujii@yokohama-cu.ac.jp }\\
  Department of Mathematical Sciences\\
  Yokohama City University\\
  Yokohama, 236--0027\\
  Japan
  }
\date{}
\maketitle
\begin{abstract}
  In this paper we treat a non--Gaussian integral and give a 
  fundamental formula in terms of discriminant. We also present 
  some related problems.
  
  This is a comment paper to arXiv:0903.2595 [math-ph] by 
  Morozov and Shakirov.
\end{abstract}
%

%
%
%     Honbun
%
%
\section{Introduction}
Gaussian plays a fundamental role in Mathematics (including 
Statistics) and Physics and et al. We want to (or we should) 
overcome its high wall in this century, and so a try is 
introduced.

In the paper \cite{MS} the following ``formula" is listed :
\begin{equation}
\label{eq:cubic-integral}
\int\int e^{-\left(ax^{3}+bx^{2}y+cxy^{2}+dy^{3}\right)}dxdy
=
\frac{1}{\sqrt[6]{-D}}
\end{equation}
where $D$ is the discriminant given by
\begin{equation}
\label{eq:cubic-discriminant}
D=-\left(27a^{2}d^{2}+4ac^{3}-18abcd-b^{2}c^{2}+4b^{3}d\right)
\end{equation}
of the cubic equation
\begin{equation}
ax^{3}+bx^{2}+cx+d=0.
\end{equation}

The equation (\ref{eq:cubic-integral}) is of course non--Gaussian. 
However, if we consider it in the framework of real category  
then (\ref{eq:cubic-integral}) is not correct because the left hand side 
diverges. In this paper we treat only real category, and so $a, b, c, d, 
x, y$ are real numbers.

Formally, by performing the change of variable \ \ $x=t\rho,\ y=\rho$\ \ 
for (\ref{eq:cubic-integral}) we have
\begin{eqnarray*}
\mbox{LHS of (\ref{eq:cubic-integral})}
&=&
\int\int e^{-\rho^{3}\left(at^{3}+bt^{2}+ct+d\right)}|\rho|dtd\rho \\
&=&
\int
\left\{\int e^{-\left(at^{3}+bt^{2}+ct+d\right)\rho^{3}}|\rho|d\rho\right\}dt 
\\
&=&
\int |\sigma|e^{-\sigma^{3}}d\sigma
\int \frac{1}{\sqrt[3]{(at^{3}+bt^{2}+ct+d)^{2}}}dt
\end{eqnarray*}
by the change of variable \ $\sigma=\sqrt[3]{at^{3}+bt^{2}+ct+d}\ \rho$. 

Therefore we can conjecture that the formula may be
\begin{equation}
\label{eq:real cubic-integral}
\int_{{\bf R}} \frac{1}{\sqrt[3]{(ax^{3}+bx^{2}+cx+d)^{2}}}dx
=
\frac{C}{\sqrt[6]{-D}}
\end{equation}
under the change $t\rightarrow x$. Here $C$ is a constant.

In the paper we calculate the left hand side of (\ref{eq:real cubic-integral}) 
{\bf directly}.

\section{Fundamental Formula}
Before stating the result let us make some preparations.
The Gamma--function $\Gamma(p)$ is defined by
\begin{equation}
\label{eq:Gamma-function}
\Gamma(p)=\int_{0}^{\infty}e^{-x}x^{p-1}dx\quad (p>0)
\end{equation}
and the Beta--function $B(p,q)$ is
\begin{equation}
\label{eq:Beta-function}
B(p,q)=\int_{0}^{1}x^{p-1}(1-x)^{q-1}dx\quad (p,\ q>0).
\end{equation}
Note that the Beta--function is rewritten as
\[
B(p,q)=\int_{0}^{\infty}\frac{x^{p-1}}{(1+x)^{p+q}}dx.
\]
See \cite{WW} in more detail. Now we are in a position to 
state the result.

\vspace{5mm}
\begin{Large}
\noindent
{\bf Integral Formula}\\
\end{Large}
(I)\ For $D < 0$ 
\begin{equation}
\label{eq:formula-I}
\int_{{\bf R}} \frac{1}{\sqrt[3]{(ax^{3}+bx^{2}+cx+d)^{2}}}dx
=
\frac{C_{-}}{\sqrt[6]{-D}}
\end{equation}
where
\[
C_{-}=\sqrt[3]{2}B(\frac{1}{2},\frac{1}{6}).
\]

\vspace{5mm}
\noindent
(II)\ For $D > 0$ 
\begin{equation}
\label{eq:formula-II}
\int_{{\bf R}} \frac{1}{\sqrt[3]{(ax^{3}+bx^{2}+cx+d)^{2}}}dx
=
\frac{C_{+}}{\sqrt[6]{D}}
\end{equation}
where
\[
C_{+}=3B(\frac{1}{3},\frac{1}{3}).
\]

\vspace{5mm}
\noindent
(III)\ $C_{-}$ and $C_{+}$ are related to $C_{+}=\sqrt{3}C_{-}$ 
by the equation
\begin{equation}
\label{eq:formula-III}
\sqrt{3}B(\frac{1}{3},\frac{1}{3})=\sqrt[3]{2}B(\frac{1}{2},\frac{1}{6}).
\end{equation}

\vspace{3mm}
Our result shows that the integral depends on the sign of $D$, and so 
our question is as follows.

\vspace{3mm}
\noindent
{\bf Problem}\ \ Can the result be derived from the method developed in 
\cite{MS} ?

\vspace{3mm}
A comment is in order. \ If we treat the Gaussian case 
(: $e^{-(ax^{2}+bxy+cy^{2})}$) then the integral is reduced to
\begin{equation}
\label{eq:gaussian-formula}
\int_{{\bf R}}\frac{1}{ax^{2}+bx+c}dx=\frac{2\pi}{\sqrt{-D}}
\end{equation}
if $a>0$ and $D=b^{2}-4ac<0$. Noting
\[
\pi=\frac{\sqrt{\pi}\sqrt{\pi}}{1}=
\frac{\Gamma(\frac{1}{2})\Gamma(\frac{1}{2})}{\Gamma(1)}=
B(\frac{1}{2},\frac{1}{2})
\]
(\ref{eq:gaussian-formula}) should be read as 
\[
\int_{{\bf R}}\frac{1}{ax^{2}+bx+c}dx=
\frac{2B(\frac{1}{2},\frac{1}{2})}{\sqrt{-D}}.
\]

\section{Discriminant}
In this section we make some comments on the discriminant 
(\ref{eq:cubic-discriminant}). See \cite{Sa} in more detail 
(\cite{Sa} is strongly recommended). 

For the equations
\begin{equation}
\label{eq:cubic-function}
f(x)=ax^{3}+bx^{2}+cx+d,\quad
f^{\prime}(x)=3ax^{2}+2bx+c
\end{equation}
the resultant $R(f,f^{\prime})$ of $f$ and $f^{\prime}$ is given by
\begin{equation}
\label{eq:resultant}
R(f,f^{\prime})=
\left|
  \begin{array}{ccccc}
   a  & b  & c  & d  & 0  \\
   0  & a  & b  & c  & d  \\
   3a & 2b & c  & 0  & 0  \\
   0  & 3a & 2b & c  & 0  \\
   0  & 0  & 3a & 2b & c 
  \end{array}
\right|.
\end{equation}
It is easy to calculate (\ref{eq:resultant}) and the result becomes
\begin{equation}
\label{eq:}
\frac{1}{a}R(f,f^{\prime})
=27a^{2}d^{2}+4ac^{3}-18abcd-b^{2}c^{2}+4b^{3}d
=-D.
\end{equation}

On the other hand, if $\alpha$, $\beta$, $\gamma$ are three solutions 
of $f(x)=0$ in (\ref{eq:cubic-function}), then the following relations 
are well--known.
\begin{equation}
\left\{
\begin{array}{ll}
\alpha +\beta +\gamma =-\frac{b}{a}                 \\
\alpha\beta +\alpha\gamma +\beta\gamma =\frac{c}{a} \\
\alpha\beta\gamma =-\frac{d}{a}
\end{array}
\right.
\end{equation}

If we set
\begin{equation}
\label{eq:Delta}
\Delta =(\alpha-\beta)(\alpha-\gamma)(\beta-\gamma)
\end{equation}
the discriminant $D$ is given by
\begin{equation}
\label{eq:definition}
D=a^{4}\Delta^{2}.
\end{equation}

\noindent
Let us calculate $\Delta^{2}$ directly. For the Vandermonde matrix
\[
V=
\left(
  \begin{array}{ccc}
   1          & 1         & 1           \\
   \alpha     & \beta     & \gamma      \\
   \alpha^{2} & \beta^{2} & \gamma^{2}   
  \end{array}
\right)\ 
\Longrightarrow\  
|V|=-\Delta 
\]
we obtain
\begin{eqnarray}
\Delta^{2}
&=&(-|V|)^{2}=|V||V^{T}|=|VV^{T}| \nonumber \\
&=&
\left|
  \begin{array}{ccc}
   3 & \alpha +\beta +\gamma & \alpha^{2}+\beta^{2}+\gamma^{2} \\
   \alpha +\beta +\gamma & \alpha^{2}+\beta^{2}+\gamma^{2} & 
   \alpha^{3}+\beta^{3}+\gamma^{3}  \\
   \alpha^{2}+\beta^{2}+\gamma^{2} & \alpha^{3}+\beta^{3}+\gamma^{3} & 
   \alpha^{4}+\beta^{4}+\gamma^{4}   
  \end{array}
\right|     \nonumber \\
&=& \cdots  \nonumber \\
&=&\frac{1}{a^{4}}
\frac{-1}{3}\left\{(bc-9ad)^{2}-4(b^{2}-3ac)(c^{2}-3bd)\right\}.
\end{eqnarray}

This result is very suggestive. In fact, from the cubic equation
\[
ax^{3}+bx^{2}+cx+d=0
\]
we have three data
\[
A=b^{2}-3ac,\quad B=bc-9ad,\quad C=c^{2}-3bd
\]
, and so if we consider the quadratic equation
\[
AX^{2}+BX+C=0
\]
then the discriminant is just $B^{2}-4AC$. This is very interesting.

\vspace{3mm}
\noindent
{\bf Problem}\ \ Make the meaning clear !

\section{Concluding Remarks}
In the paper we calculated the non--Gaussian integral 
(\ref{eq:real cubic-integral}) in a direct manner.  
Details of calculation will be published in \cite{Fu}. 

In this stage we can consider the general case. For the general 
equation
\begin{equation}
\label{eq:general-equation}
f(x)=a_{0}x^{n}+a_{1}x^{n-1}+\cdots +a_{n-1}x+a_{n}
\end{equation}
the (non--Gaussian) integral becomes
\begin{equation}
\label{eq:general-integral}
\int_{{\bf R}}\frac{1}{\sqrt[n]{f(x)^{2}}}dx.
\end{equation}

The discriminant $D$ of the equation $f(x)=0$ is given by 
the resultant $R(f,f^{\prime})$ of $f$ and $f^{\prime}$ like
\begin{equation}
\frac{1}{a_{0}}R(f,f^{\prime})=(-1)^{\frac{n(n-1)}{2}}D
\ \Longleftrightarrow \ 
D=(-1)^{\frac{n(n-1)}{2}}R(f,f^{\prime})/a_{0}
\end{equation}
where
\[
R(f,f^{\prime})=
\left|
  \begin{array}{cccccccc}
   a_{0} & a_{1} & \cdots & a_{n-1} & a_{n} &  &  &       \\
      & a_{0} & a_{1} & \cdots & a_{n-1} & a_{n} &  &     \\
      &  & \ddots &    & & \ddots  &   &                  \\
      &  & & a_{0} & a_{1} & \cdots & a_{n-1} & a_{n}     \\
   na_{0} & (n-1)a_{1} & \cdots & a_{n-1} &    & & &      \\
      & na_{0} & (n-1)a_{1} & \cdots & a_{n-1} & & &      \\
      &  & \ddots &   &   & \ddots   &   &                \\      
      & & & & na_{0} & (n-1)a_{1} & \cdots & a_{n-1}        
  \end{array}
\right|,
\]
see (\ref{eq:cubic-function}) and (\ref{eq:resultant}).

\vspace{5mm}
For example, if $n=4$ and $n=5$ then we have
\begin{eqnarray*}
D_{n=4}
&=&256a_{0}^{3}a_{4}^{3}-4a_{1}^{3}a_{3}^{3}
-27a_{0}^{2}a_{3}^{4}-27a_{1}^{4}a_{4}^{2}
-128a_{0}^{2}a_{2}^{2}a_{4}^{2}
+a_{1}^{2}a_{2}^{2}a_{3}^{2}+16a_{0}a_{2}^{4}a_{4}   \\
&{}&
-4a_{0}a_{2}^{3}a_{3}^{2}-4a_{1}^{2}a_{2}^{3}a_{4}
+144a_{0}^{2}a_{2}a_{3}^{2}a_{4}-6a_{0}a_{1}^{2}a_{3}^{2}a_{4}
+144a_{0}a_{1}^{2}a_{2}a_{4}^{2}
-192a_{0}^{2}a_{1}a_{3}a_{4}^{2}  \\
&{}&
+18a_{0}a_{1}a_{2}a_{3}^{3}
+18a_{1}^{3}a_{2}a_{3}a_{4}-80a_{0}a_{1}a_{2}^{2}a_{3}a_{4},
\end{eqnarray*}
and

\begin{eqnarray*}
D_{n=5}
&=&3125a_{0}^{4}a_{5}^{4}-2500a_{0}^{3}a_{1}a_{4}a_{5}^{3}
-3750a_{0}^{3}a_{2}a_{3}a_{5}^{3}+2000a_{0}^{3}a_{2}a_{4}^{2}a_{5}^{2}
+2250a_{0}^{3}a_{3}^{2}a_{4}a_{5}^{2} \\
&{}&
-1600a_{0}^{3}a_{3}a_{4}^{3}a_{5}+256a_{0}^{3}a_{4}^{5}
+2000a_{0}^{2}a_{1}^{2}a_{3}a_{5}^{3}-50a_{0}^{2}a_{1}^{2}a_{4}^{2}a_{5}^{2}
+2250a_{0}^{2}a_{1}a_{2}^{2}a_{5}^{3} \\
&{}&
-2050a_{0}^{2}a_{1}a_{2}a_{3}a_{4}a_{5}^{2}+160a_{0}^{2}a_{1}a_{2}a_{4}^{3}a_{5}
-900a_{0}^{2}a_{1}a_{3}^{3}a_{5}^{2}+1020a_{0}^{2}a_{1}a_{3}^{2}a_{4}^{2}a_{5}
-192a_{0}^{2}a_{1}a_{3}a_{4}^{4} \\
&{}&
-900a_{0}^{2}a_{2}^{3}a_{4}a_{5}^{2}+825a_{0}^{2}a_{2}^{2}a_{3}^{2}a_{5}^{2}
+560a_{0}^{2}a_{2}^{2}a_{3}a_{4}^{2}a_{5}-128a_{0}^{2}a_{2}^{2}a_{4}^{4}
-630a_{0}^{2}a_{2}a_{3}^{3}a_{4}a_{5} \\
&{}&
+144a_{0}^{2}a_{2}a_{3}^{2}a_{4}^{3}+108a_{0}^{2}a_{3}^{5}a_{5}
-27a_{0}^{2}a_{3}^{4}a_{4}^{2}-1600a_{0}a_{1}^{3}a_{2}a_{5}^{3}
+160a_{0}a_{1}^{3}a_{3}a_{4}a_{5}^{2} \\
&{}&
-36a_{0}a_{1}^{3}a_{4}^{3}a_{5}+1020a_{0}a_{1}^{2}a_{2}^{2}a_{4}a_{5}^{2}
+560a_{0}a_{1}^{2}a_{2}a_{3}^{2}a_{5}^{2}-746a_{0}a_{1}^{2}a_{2}a_{3}a_{4}^{2}a_{5}
+144a_{0}a_{1}^{2}a_{2}a_{4}^{4} \\
&{}&
+24a_{0}a_{1}^{2}a_{3}^{3}a_{4}a_{5}-6a_{0}a_{1}^{2}a_{3}^{2}a_{4}^{3}
-630a_{0}a_{1}a_{2}^{3}a_{3}a_{5}^{2}+24a_{0}a_{1}a_{2}^{3}a_{4}^{2}a_{5}
+356a_{0}a_{1}a_{2}^{2}a_{3}^{2}a_{4}a_{5} \\
&{}&
-80a_{0}a_{1}a_{2}^{2}a_{3}a_{4}^{3}-72a_{0}a_{1}a_{2}a_{3}^{4}a_{5}
+18a_{0}a_{1}a_{2}a_{3}^{3}a_{4}^{2}+108a_{0}a_{2}^{5}a_{5}^{2}
-72a_{0}a_{2}^{4}a_{3}a_{4}a_{5}+16a_{0}a_{2}^{4}a_{4}^{3} \\
&{}&
+16a_{0}a_{2}^{3}a_{3}^{3}a_{5}-4a_{0}a_{2}^{3}a_{3}^{2}a_{4}^{2}
+256a_{1}^{5}a_{5}^{3}-192a_{1}^{4}a_{2}a_{4}a_{5}^{2}
-128a_{1}^{4}a_{3}^{2}a_{5}^{2}+144a_{1}^{4}a_{3}a_{4}^{2}a_{5} \\
&{}&
-27a_{1}^{4}a_{4}^{4}+144a_{1}^{3}a_{2}^{2}a_{3}a_{5}^{2}
-6a_{1}^{3}a_{2}^{2}a_{4}^{2}a_{5}-80a_{1}^{3}a_{2}a_{3}^{2}a_{4}a_{5}
+18a_{1}^{3}a_{2}a_{3}a_{4}^{3}+16a_{1}^{3}a_{3}^{4}a_{5} \\
&{}&
-4a_{1}^{3}a_{3}^{3}a_{4}^{2}-27a_{1}^{2}a_{2}^{4}a_{5}^{2}
+18a_{1}^{2}a_{2}^{3}a_{3}a_{4}a_{5}-4a_{1}^{2}a_{2}^{3}a_{4}^{3}
-4a_{1}^{2}a_{2}^{2}a_{3}^{3}a_{5}+a_{1}^{2}a_{2}^{2}a_{3}^{2}a_{4}^{2}.
\end{eqnarray*}
However, to write down the general case explicitly is not easy.

\vspace{5mm}
\noindent
{\bf Problem}\ \ Calculate (\ref{eq:general-integral}) for $n=4$ (and 
$n=5$) directly.

\vspace{5mm}
The wall called Gaussian is very high and not easy to overcome, 
and therefore hard work will be needed.

%%%%%%%%%%%%%
%References%
%%%%%%%%%%%%%


\begin{thebibliography}{99}
%
\bibitem{MS}A. Morozov and Sh. Shakirov :
\newblock Introduction to Integral Discriminants, 
\newblock arXiv:0903.2595 [math-ph].
%
\bibitem{WW}E. T. Whittaker and G. N. Watson :
\newblock A Course of MODERN ANALYSIS, 
\newblock 1990 (latest), Cambridge University Press.
%
\bibitem{Sa}I. Satake : 
\newblock Linear Algebra (in Japanese), 
\newblock 1989 (latest), Shokabo, Tokyo. \\
\newblock As far as I know this is the best book on Elementary 
Linear Algebra.
%
\bibitem{Fu}K. Fujii : 
\newblock in preparation. 
%
\end{thebibliography}
\end{document}